\documentclass[a4paper]{jpconf}
\usepackage{graphicx}
\usepackage{amsmath}
\usepackage{lineno}

\begin{document}
\title{Inclusive Flavour Tagging Algorithm}

\author{Tatiana Likhomanenko$^{1,2,3}$, Denis Derkach$^{1, 2}$, Alex Rogozhnikov$^{1,2}$}


\address{$^1$ National Research University Higher School of Economics (HSE), RU}
\address{$^2$ Yandex School of Data Analysis (YSDA), RU}
\address{$^3$ NRC "Kurchatov Institute", RU}

\ead{tata.antares@yandex.ru}

\begin{abstract}
Identifying the flavour of neutral $B$ mesons production is one of the most important components needed in the study of time-dependent 
$CP$ violation.
The harsh environment of the Large Hadron Collider makes it particularly hard to succeed in this
task. 
We present an inclusive flavour-tagging algorithm as an upgrade of the algorithms currently used by the LHCb
experiment.
Specifically, a probabilistic model which efficiently combines information from reconstructed
vertices and tracks using machine learning is proposed. 
The algorithm does not use information about underlying physics process. It reduces the dependence
on the performance of lower level identification capacities and thus increases the overall
performance.
The proposed inclusive flavour-tagging algorithm is applicable to tag the flavour of $B$ mesons in
any proton-proton experiment.
\end{abstract}

\section{Introduction}
$B$ mesons contain either a $b$ or a $\bar{b}$ quark, which defines their flavour. 
The flavour-tagging (FT) algorithms determine the flavour of a reconstructed signal $B$ meson candidate at
the production point in proton-proton collisions. 
The FT algorithms are used to measure differences in the behaviour of particles and antiparticles
(e.g.\ measurements of flavour oscillations of $B_{(s)}^0$ mesons) and $CP$ asymmetries to probe the
validity of the Standard Model of particle physics. 

\begin{figure}
	\begin{center}
		\begin{minipage}{22pc}
			\includegraphics[width=22pc]{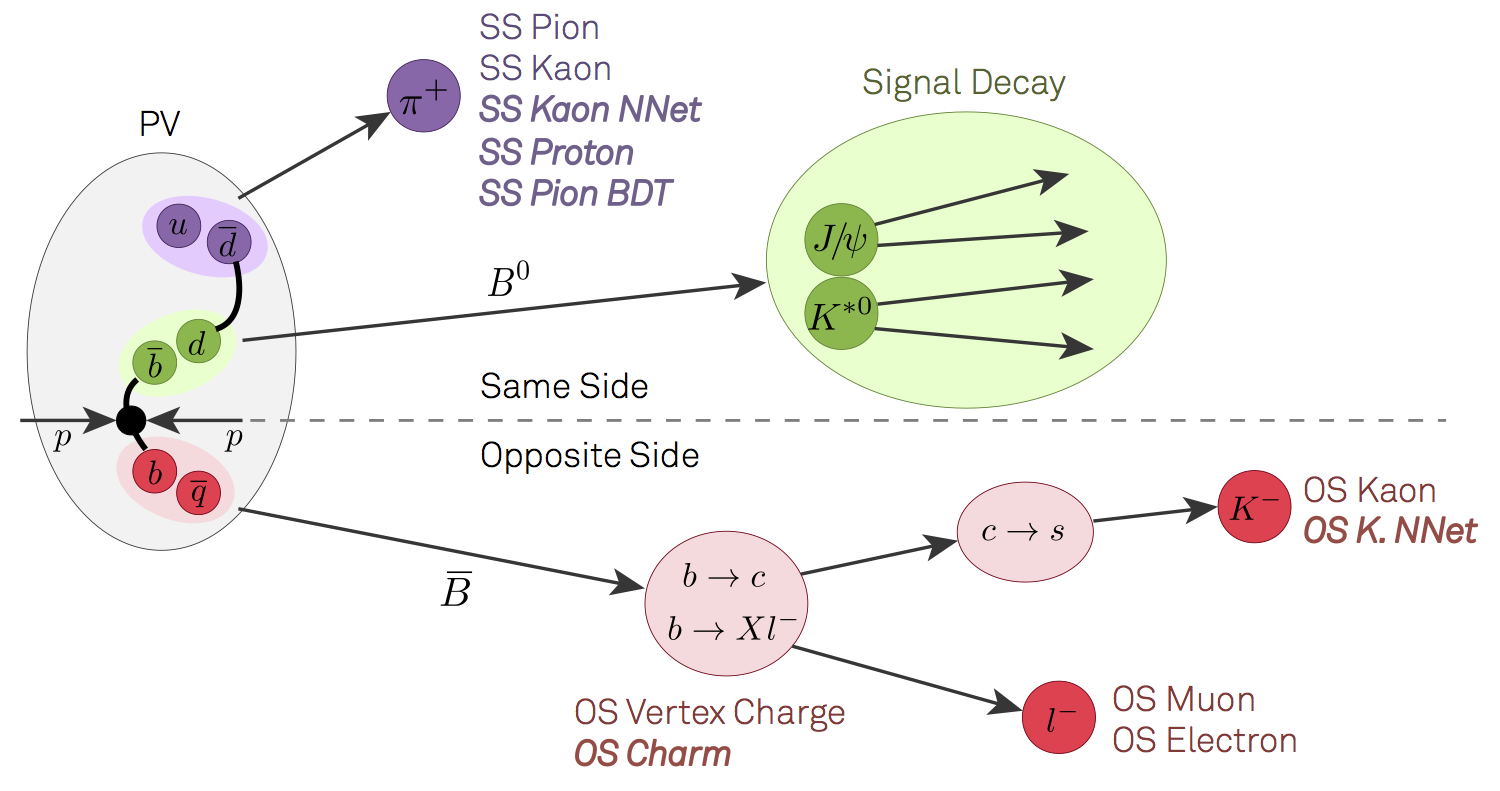}
		\end{minipage}
    \vspace{-0.5cm}      
    \caption{\label{scheme} Schematic view of the different sources of information available to
             define the initial flavour of a signal $B$ candidate.
            }
	\end{center}
  \vspace{-1cm}
\end{figure}

The production of a $B$ meson is usually accompanied by the production of another $b$ hadron and
other particles like kaons, pions, and protons (see Figure~\ref{scheme}). 
At hadron collider experiments the FT algorithms are usually divided into two groups:

\begin{itemize}
  \item opposite side (OS) taggers use the decay products of $b$ hadrons that are produced together
    with the signal $B$ (see~\cite{lhcbOS});
  \item same side (SS) taggers exploit light particles that evolve from the hadronisation process of
    the signal $B$ meson like kaons, pions, and protons (see~\cite{lhcbSS}).
\end{itemize}

The current version of the FT algorithm used by the LHCb~\cite{lhcbOS, lhcbSS, lhcbcharm},
CMS~\cite{cms}, Atlas~\cite{atlas}, CDF~\cite{cdf} and D0~\cite{d0} experiments tries to identify tracks and vertices produced on the OS/SS sides (SS tagging is done only by LHCb and CDF). It works
as follows:

\begin{enumerate}
  \item the first step finds all tagging tracks and tagging vertices, where the latter is only used
    in OS taggers\footnote{Note, a tagging track/vertex is a track/vertex whose charge is used to
    predict the flavour of the signal $B$. Ideal choices for a tagging track are pion, kaon
    or proton tracks involved in the SS tagger and lepton or kaon tracks coming from a $b$ hadron decay in the
    OS tagger. Ideal choice of a tagging vertex for the OS is a $b$ hadron decay vertex or charm hadron vertex coming from a $b$ hadron decay.}.
    Both the OS and SS algorithms find one, maybe several, tagging track, while the OS algorithm also finds one tagging
    vertex. 

	\begin{enumerate}
		\item for the OS algorithm only lepton, kaon tracks and $b$ hadron decay, secondary charm hadron vertices are considered;
		\item for the SS algorithm only pion, proton or kaon tracks are considered;
    \item other physically motivated selections are applied to leave only tracks/vertices which have the
      characteristics that help to define the flavour;
    \item if more than one track (vertex) is left after the previous steps for the OS or SS, a
      special rule is applied to select an appropriate track (vertex);
	\end{enumerate}

  \item each of the OS/SS algorithms predicts a flavour based on the charge of the tagging track
    (vertex). Other characteristics of the track (vertex) are used to estimate the probability of
    incorrectly predicted flavour (i.e.\ misclassification, or mistag probability).

  \item finally, the predictions of the OS and SS taggers are usually combined in the flavour-tagged
    analyses.
\end{enumerate}

The first step naturally follows our physics intuition, but requires setting ad-hoc
conditions, which require a deep understanding of the physics 
processes. From an analysis point of view, this pipeline causes some disadvantages:
\begin{itemize}
  \item the algorithm relies heavily on the particle identification and reconstructed variables
    during the selection of the tagging tracks (vertices);
  \item the process of selecting a tagging particle is based on physics assumptions. This prevents the use of complex selection rules;
	\item a lot of information is lost since only a couple of tracks (vertices) are selected.
\end{itemize}

This paper describes a new approach to define the signal $B$ flavour that exploits all available
information in an event without using information about the underlying physics processes, like a
tagging track (vertex) search.

\section{Inclusive Probabilistic Model}

The algorithm starts using an inclusive probabilistic model, which combines information from all
tracks and vertices for each selected event containing a $B$ candidate to tag. It uses an assumption similar to a naive Bayes
model.
Specifically, it assumes a strong independence of the tagging information available in the tracks 
and vertices\footnote{Note that the use of a varying number of multipliers for each event is atypical
for a naive Bayes approach in machine learning.}.

Let ``components'' refer to both tracks and vertices. Additionally, let $s_p$ be the charge sign of a
component ($+1$ or $-1$) and $s_b$ be the flavour of the signal $B$ ($+1$ for $\bar{b}$ and $-1$ for 
$b$). Then, assume the following:
\[
\frac{P(\bar{b})}{P(b)} = \prod_{\text{components}} \frac{P(\bar{b} | B, \text{component}, s_p)}{P(b | B, \text{component}, s_p)} = 
\prod_{\text{components}} \left(
	\frac{P(s_b \cdot s_p > 0| B, \text{component})}{P(s_b \cdot s_p < 0| B, \text{component})}
\right)^{s_p}
\]
The last equality assumes that the spurious asymmetries introduced by different detection
efficiencies for particles and antiparticles in the different regions of the detectors are
negligible.

The usage of this formula, however, requires estimating probabilities ${P(s_b \cdot s_p > 0| B,
 \text{component}})$ and $P(s_b \cdot s_p < 0| B, \text{component}).$ Note that these probabilities are established using different parameters of the signal $B$ meson and a track/vertex, but not
 using their charges.

This approach has several key properties:
\begin{itemize}
  \item it combines all available information from the components of the events under the naive
    probabilistic model;
  \item it implicitly determines the tagging tracks and vertices by the value of the ratio of the 
    probabilities. Most of the particles will have a very small contribution;
  \item it does not depend on the tagging particle type (i.e.\ pion, kaon, electron, muon, proton)
    and it is not split into OS and SS tagging algorithms;
	\item it is symmetric with respect to matter/antimatter due to model definition.
\end{itemize}
Thus, the proposed FT algorithm is an inclusive model.



\section{Inclusive Training}


Charged $B$ meson can be tagged using the charge of its decay components. 
Thus, the flavour of the meson can be defined ($\bar{b}$ for $B^+$ and $b$ for
$B^-$). 
${B^{\pm} \to J/\psi [\mu^+\mu^-]K^{\pm}}$ decays are used for training. 
The charge of the kaon in the signal decay is used to independently infer the flavour of the $B$
meson at production: ${P(\bar{b}) = P(B^+),} \, {P(b) = P(B^-)}$. The inclusive model is applied to the
LHCb data samples that contain reconstructed signal decays $B^{\pm} \to J/\psi [\mu^+\mu^-]
K^{\pm}$. 
The set of all tracks with the low probability to be ghost (fake track) and vertices for all events form the tracks and vertices datasets. Note the
tracks and vertices forming the reconstructed signal decay are excluded.

In the probabilistic model conditional probabilities ${P(s_b \cdot s_p > 0| B, \text{component})}$ and ${P(s_b \cdot s_p < 0| B, \text{component})}$ are unknown. We can estimate them using a classification
model. The target for this classification model is:
\[
	\text{target} = 
	\begin{cases}
		1, & \text{ if } s_b \cdot s_p > 0, \\
		0, & \text{ if } s_b \cdot s_p < 0.
	\end{cases}
\]

Two gradient boosted decision tree (GBDT) algorithms are trained to predict the conditional
probability $P(s_b \cdot s_p > 0| B, \text{component})$ for tracks and vertices. 
Kinematic properties of the tracks, vertices and signal $B$ meson, information from the
particle identification algorithm based on machine learning methods and track quality criteria are used as input observables. 
For the $B$ meson the following features are used: transverse momentum, polar angle, impact
parameter with respect to the primary interaction, pseudorapidity. For tracks, the particle identification algorithm output, polar angle, momentum,
transverse momentum are used. Finally, for vertices, the number of tracks forming the vertex, mean of tracks impact
parameters and mean of their transverse momenta, mass and momentum, which are calculated assuming pion mass for the incoming tracks, lifetime, angle between the signal $B$ and
the vertex are used.

\section{Symmetric Calibration}

The conditional probability $P(s_b \cdot s_p > 0| B, \text{component})$ predicted by a
classification model (i.e.\ by the GBDT) may be biased (see~\cite{calib},~\cite{calib_bdt}).
Additionally, $P(\bar{b})$ and $P(b)$ computed by the probabilistic model may not be true
probabilities due to the naive Bayes assumption. 
To compensate for these biases, the classifier output must be calibrated.
Furthermore, the model should have the same behaviour for particles and antiparticles
except small asymmetry of the production and detectors. 
This means that distributions for $P(B^+)$ and $P(B^-)$ should be symmetric around $0.5$.

\begin{figure}
	\begin{center}
		\begin{minipage}{30pc}
			\begin{center}
				\includegraphics[width=24pc]{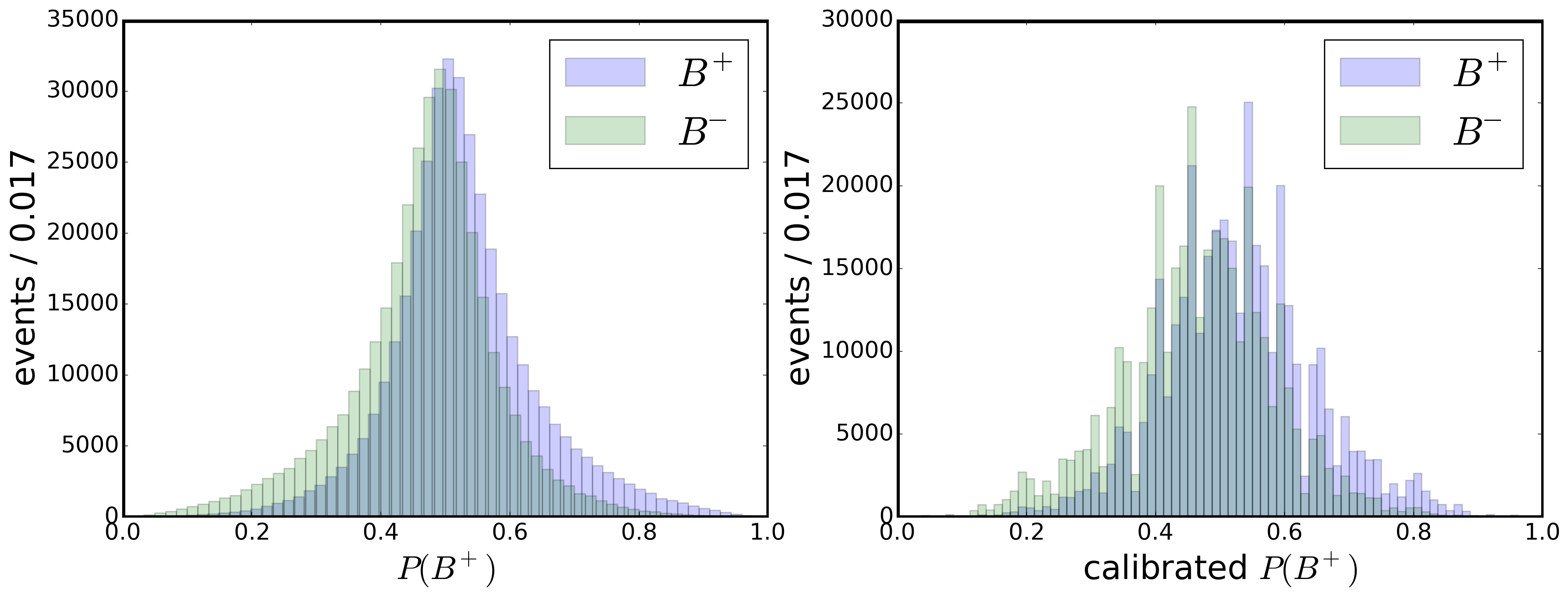}
        \vspace{-0.4cm}      
        \caption{\label{fig:pdf} $P(B^+)$ distribution for $B^+$ (blue) and $B^-$ (green) before calibration (left)
                 and after isotonic regression calibration (right).}
			\end{center}
		\end{minipage}
	\end{center}
  \vspace{-0.7cm}
\end{figure}

\begin{figure}

	\begin{center}
		\begin{minipage}{17pc}
			\hspace{1.5cm}\includegraphics[width=13pc]{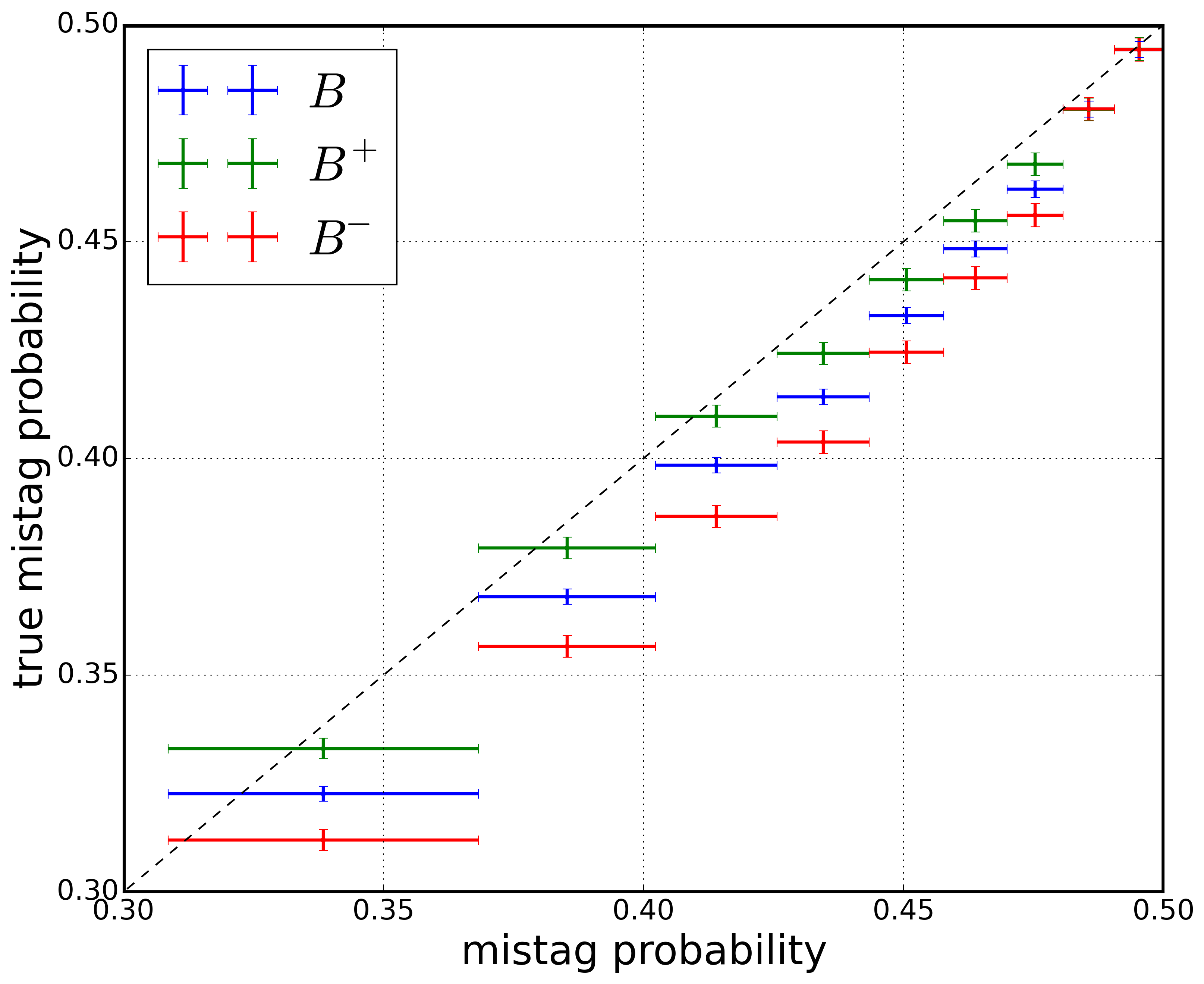}
		\end{minipage}
		\begin{minipage}{17pc}
			\includegraphics[width=13pc]{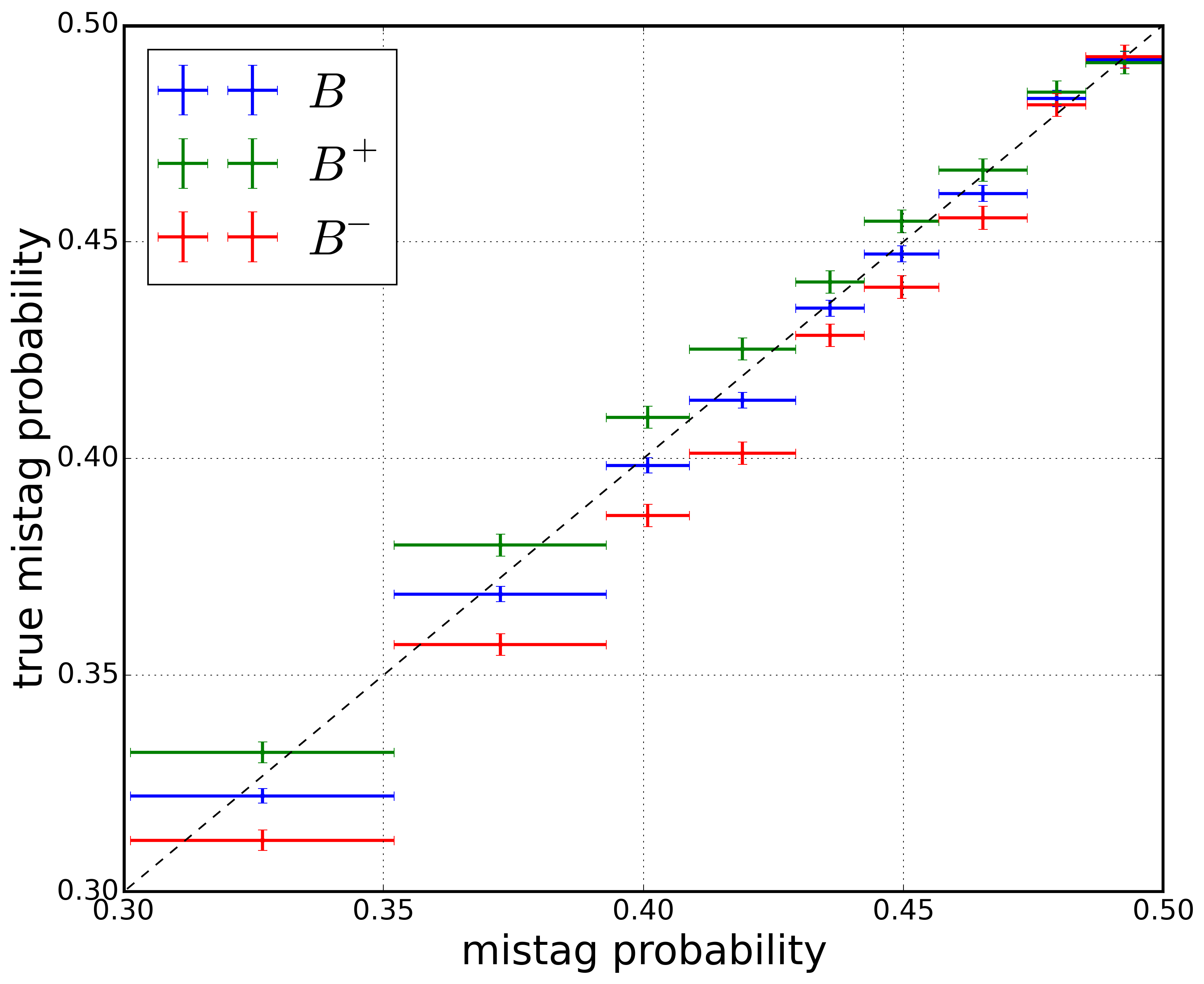}
		\end{minipage} 
    \vspace{-0.4cm}      
    \caption{\label{fig:calib} Calibration and symmetry check before (left) and after (right)
            isotonic regression. Estimation of the frequency based probability is used as a true
            probability in each bin for percentile 10 binning.}
	\end{center}
  \vspace{-0.9cm}
\end{figure}

To calibrate the GBDT output, Platt scaling \cite{platt} and isotonic regression \cite{isotonic} were
used. Platt scaling provides better results than isotonic regression and the uncalibrated probabilities. 
When calibrating $P(B^+)$ and $P(B^-)$, symmetric isotonic regression is used to preserve 
symmetry in the distributions. 
The calibration rule, $f$, is required to be symmetric, i.e.\ $f(1 - x) = 1 - f(x)$, where $x$ is
$P(B^+)$.
Figure~\ref{fig:pdf} shows distributions for $B^+$ and $B^-$ before and after the isotonic
regression calibration. The comparison between the probability obtained from the inclusive model and
frequency based estimation of the true probability is shown in Figure~\ref{fig:calib} before and
after the calibration procedure. 
Distributions for $P(B^+)$ and $P(B^-)$ are checked to be symmetric around $0.5$ after calibration (see Figure \ref{fig:symmetry}).
After the calibration the inclusive model has improved Brier and logarithmic scores (see the scoring
rules \cite{scoring}), while the Platt scaling gives worse scores than isotonic regression.

\section{Quality Metric}

The figure of merit of a FT algorithm is the effective efficiency (see \cite{lhcbOS, lhcbSS, lhcbcharm})
since the overall statistical power of the flavour-tagged sample is proportional to it.
As a proxy metric of the effective efficiency the ROC curve is used in the analysis to optimize the
FT algorithm. After the effective efficiency is checked to have increased value with respect to the previous results. 

\begin{figure}
	\begin{center}
		\begin{minipage}{18pc}
      \begin{center}
			\includegraphics[width=13pc]{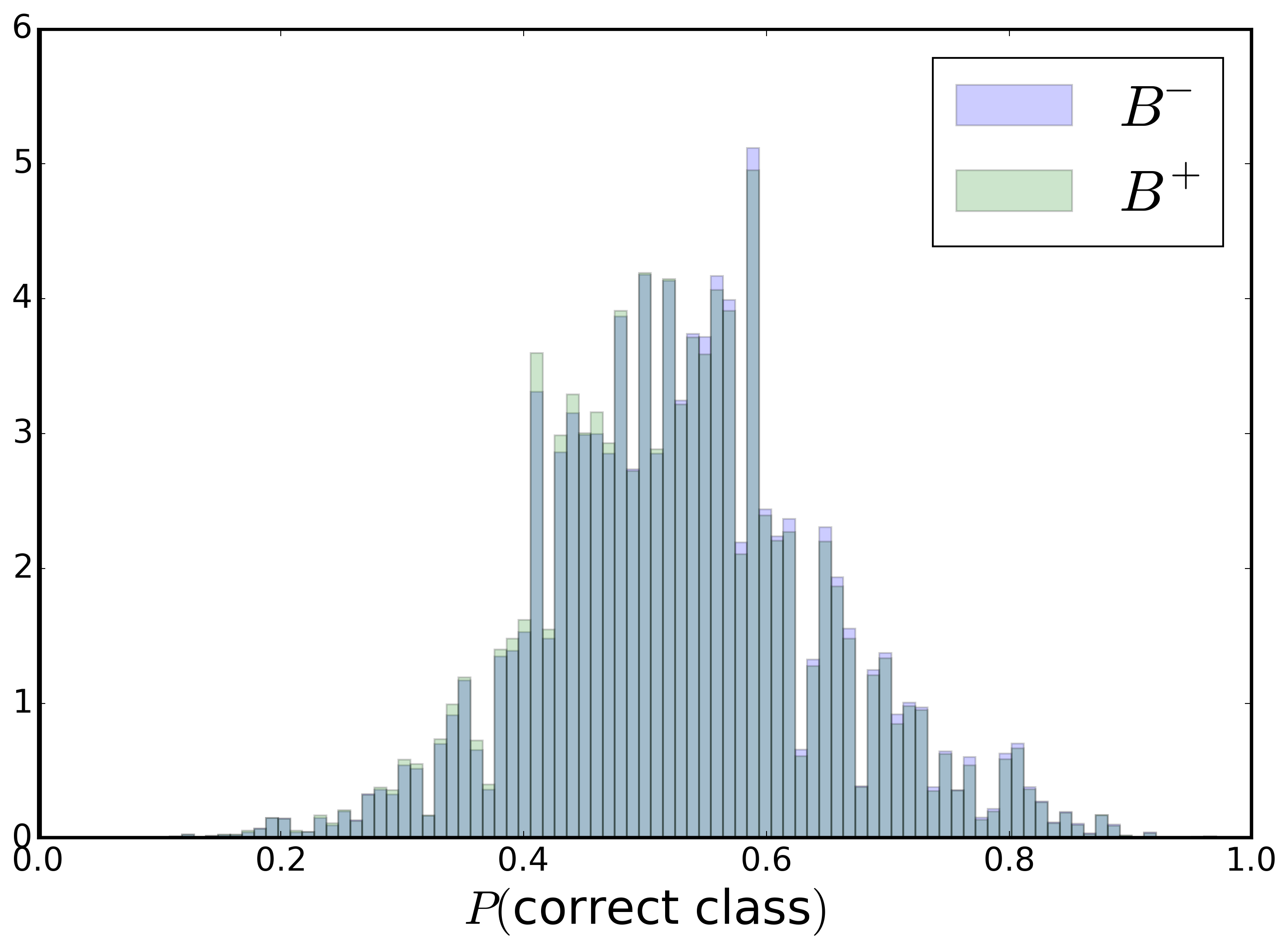}
      \vspace{-0.5cm}
			\caption{\label{fig:symmetry} $P(B^+)$ for $B^+$ (blue) and $P(B^-)$ for $B^-$ (green) distributions. Kolmogorov-Smirnov distance is $0.0163$.} 
      \end{center}
		\end{minipage}
		\hspace{1cm}
		\begin{minipage}{17pc}
			\begin{center}
				\includegraphics[width=12pc]{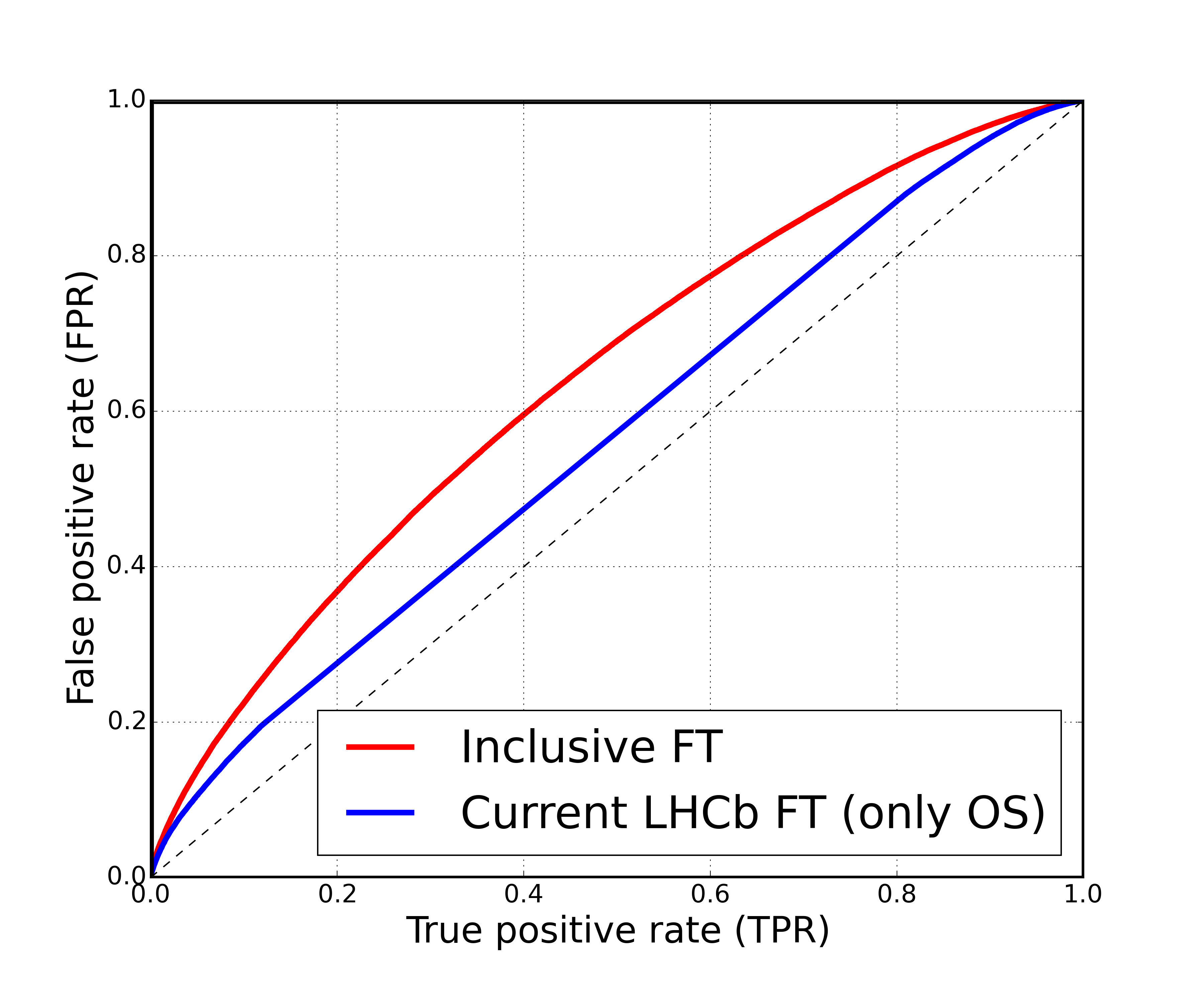}
        \vspace{-0.5cm}
				\caption{\label{fig:roc} ROC curves for the current LHCb OS FT algorithm (blue) and proposed inclusive model (red).}
			\end{center}
		\end{minipage} 
	\end{center}
\vspace{-0.9cm}
\end{figure}

We analyze the ROC curves and check that the new version has a higher ROC curve at each point.
The comparison of the
ROC curve is shown in Figure~\ref{fig:roc} and the AUC (area under the ROC curve) values are 0.566 for
the current OS FT.
The AUC for the proposed inclusive model is $0.641$.
The ROC curve and AUC values were computed for all events including untagged
events to compute the overall quality of the algorithm\footnote{Untagged events are those events for which all tracks and vertices did not pass
selections; for them probabilities are set $P(B^+) = P(B^-) = 0.5$.}.

\section{Conclusion}

We proposed a simple flavour tagging technique, which efficiently combines information from vertices
and tracks using a machine learning approach. The inclusive flavour tagging algorithm does not use
information about underlying physics process and it is applicable to FT of $B$ mesons in
proton-proton experiments. The results demonstrate significant improvement in LHCb data, as seen in the ROC AUC
score improvement from 0.566 to 0.641.

\end{document}